\documentclass{article}
\usepackage[utf8]{inputenc}

\usepackage[numbers]{natbib}
\usepackage{float}
\usepackage{comment}
\usepackage{amsmath,amsthm,amssymb,mathtools,booktabs}
\usepackage{hyperref}
\usepackage{subcaption}
\usepackage[colorinlistoftodos]{todonotes}
\usepackage[a4paper, total={6.5in, 10.in}]{geometry}
\usepackage{cleveref}
\usepackage{tabularx}
    \newcolumntype{C}{@{}>{\centering\arraybackslash}X@{}}

\theoremstyle{plain}
\newtheorem{theorem}{Theorem}[section]

\newtheorem{lemma}[theorem]{Lemma}
\newtheorem{definition}[theorem]{Definition}
\newtheorem{remark}[theorem]{Remark}

\DeclareMathOperator{\conv}{conv}
\newtheorem{Question}{Question}
\DeclareMathOperator{\T}{\mathcal{T}}
\DeclareMathOperator{\Oct}{\mathcal{O}}

\title{On inside-out Dissections of Polygons and Polyhedra}
\author{Reymond Akpanya
\thanks{RWTH Aachen University, Aachen, Germany, Reymond.akpanya@rwth-aachen.de}
\and
Adi Rivkin \thanks{Dept. of Computer Science, Technion--Israel Institute of Technology, Haifa, Israel, adi.rivkin@campus.technion.ac.il}
\and
Frederick Stock
\thanks{Miner School of Computer \& Information Sciences, Lowell, MA. USA, Frederick\_Stock@student.uml.edu} }
\date{}

\begin{document}
\maketitle
\begin{abstract}
In this work we study inside-out dissections of polygons and polyhedra.
   We first show that an arbitrary polygon can be inside-out dissected with $2n+1$ pieces, thereby improving the best previous upper bound of $4(n-2)$ pieces. Additionally, we establish that a regular polygon can be inside-out dissected with at most $6$ pieces. Lastly, we prove that any polyhedron that can be decomposed into finitely many regular tetrahedra and octahedra can be inside-out dissected.
\end{abstract}
\section{Introduction} 
In computational geometry, the term \emph{dissection} refers to the concept of decomposing (2D or 3D) shapes into smaller pieces of the same dimension, which can then be reassembled via continuous motions to form new shapes. 
Over the years the notion of dissections has been examined in various settings by formulating restrictions on the continuous motions that are allowed to rearrange the smaller pieces of a given decomposition into a new shape. For various examples of dissections we refer to \cite{dudeneydissection,demaine2005hinged,eppstein2001hingedkitemirrordissection,eppstein2023orthogonal,goldberg1964dissection,urrutia23efficient} to name only a few.

\begin{figure}[H]
    \centering
\includegraphics[width=0.5\linewidth]{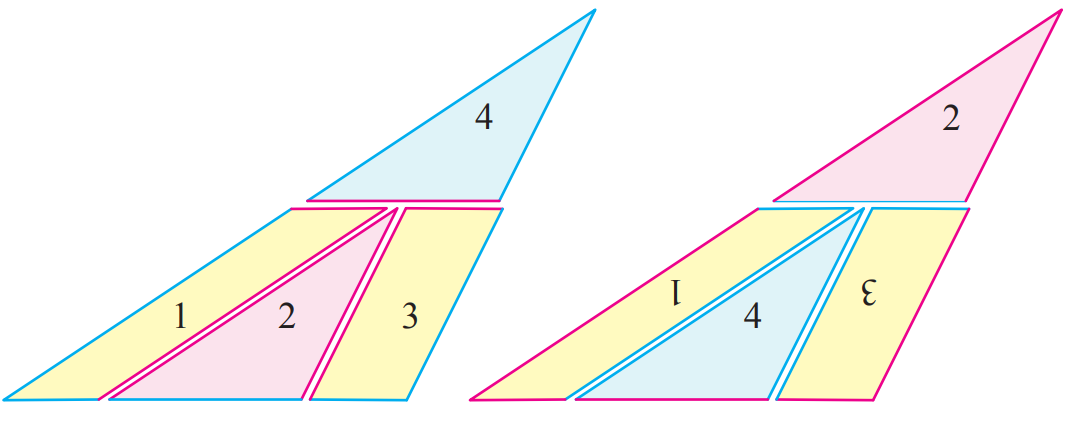}
    \caption{An inside-out-dissection of an obtuse triangle using four pieces by 
Aaron Meyerowitz.\cite{AM14}}
    \label{fig:inside-out-dissectionTriangle}
\end{figure}

In this work we study so-called \emph{inside-out dissections} of polygons and polyhedra which were introduced by Joseph O'Rourke in \cite{OR14}. Here, we recall the definition of the inside-out dissection of a polygon (polyhedron). 
\begin{definition}
Let $P$ be a polygon (polyhedron).
An \emph{\textbf{inside-out dissection}} of $P$ is a decomposition of $P$ into finitely many polygons (polyhedra) $P_1,\ldots,P_k$ such that  
\begin{enumerate}
    \item the polygons $P_1,\ldots,P_k$ can be rearranged by only applying rotations and translations to form a polygon (polyhedron)  $P'$ that is congruent to $P$,
    \item the boundary of the polygon (polyhedron) $P'$ is composed of internal cuts of $P$.
\end{enumerate}  In the case that such a decomposition of $P$ exists, we denote it by $(P_1,\ldots,P_k)$ and say that $P$ can be \emph{\textbf{inside-out dissected}}. We refer to the polygons (polyhedra) $P_1,\ldots,P_k$ as \emph{\textbf{pieces}} of $P$.
Furthermore, we define $\mathcal{I}(n)$ as the smallest natural number such that every $n$-gon $P$ can be inside-out dissected with $k\leq\mathcal{I}(n)$ pieces. 
\end{definition}
We note that, in contrast to Dudeney and hinged mirror dissections, the pieces of an inside-out dissection of a polygon (polyhedron) are not required to be connected by hinges. Figure \ref{fig:inside-out-dissectionTriangle} shows a possible inside-out dissection of an obtuse angled triangle.
The illustrated inside-out dissection of the given triangle consists of $4$ pieces. It can be shown that every triangle can be inside-out dissected with $4$ pieces, see \cite{AM14}. It is well-known that a polygon $P$ with $n$ edges can be triangulated with exactly $n-2$ triangles (proof by induction). 
Therefore by the same result the inequality $\mathcal{I}(n)\leq 4(n-2)$ holds. 
This instance yields the question whether this existing bound on the number $\mathcal{I}(n)$ can be improved.
\begin{Question}\label{Q1}
Let $P$ be an arbitrary $n$-gon. Can we show that $P$ can be inside-out dissected with $k< 4(n-2)$ pieces?
\end{Question}
Here, we show that the inequality $\mathcal{I}(n)\leq 2n+1$ is true for all $n$-gons. We also prove that if $P$ is a regular polygon, then it can be inside-out dissected with at most six pieces.
Moreover, we also investigate inside-out dissections in the three-dimensional case. In particular, we examine the following problem.
\begin{Question}\label{Q2}
Can every (or any) tetrahedron be inside-out-dissected?
\end{Question}
Both Question \ref{Q1} and Question \ref{Q2}, were posed by Joseph O'Rourke at the 36th Canadian Conference on Computational Geometry, see \cite{openproblems}.
In this paper, we answer these questions and illustrate our current results.

\section{Dissections of Polygons}
We start with the two-dimensional case and answer Question \ref{Q1}. We present a method to inside-out dissect a polygon with $n$ edges with $2n + 1$ pieces.

\begin{figure}[H]
    \centering    \includegraphics[width=\linewidth]{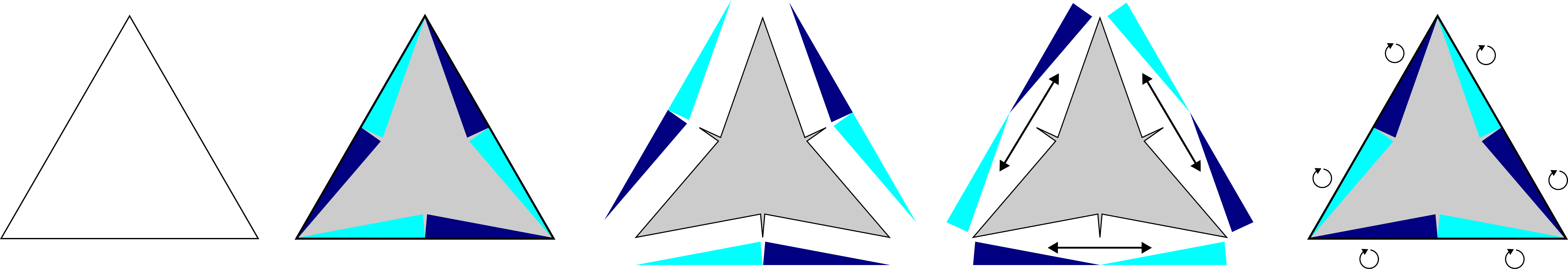}
    \caption{A triangle can be inside-out dissected with 7 pieces.}
    \label{fig:triangle}
\end{figure}
\begin{figure}[ht]
    \centering
    \includegraphics[width=.8\linewidth]{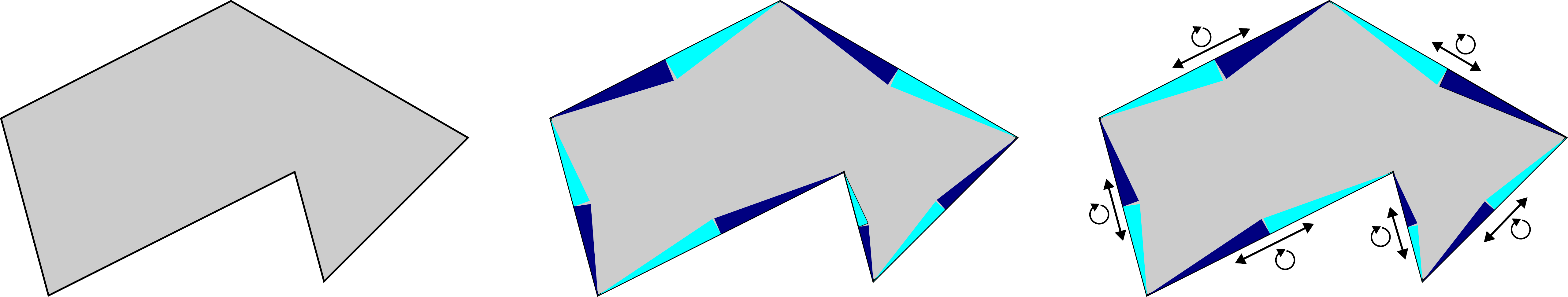}
    \caption{A generic polygon can be inside-out dissected in $2n + 1$ pieces.}
    \label{fig:adi-generic}
\end{figure}

\begin{lemma}\label{lem:triangle}
    Any $n$-gon $P$ can be inside-out dissected with $2n + 1$ pieces. In particular, this means $\mathcal{I}(n)\leq 2n+1.$
\end{lemma}
\emph{Proof Sketch.} First, note that we can inside-out-dissect a triangle $T$ with $2$ pieces per edge and one extra leftover ``scrap'' piece as follows. For every edge $e$ of $T$, construct two congruent isosceles triangles $t_1, t_2$, so their congruent edges are of length $|e|/2$. The triangles $t_1$ and $t_2$ can then be rotated $180^\circ$ and then replace each other. This rotation places the internal edge of length $|e|$ of $t_1$ and $t_2$ on the perimeter of $T$, and their external edge is now internal to $T$.
This can be done for each edge of $T$, giving $2\cdot 3$ pieces and one for $2\cdot3 + 1 = 7$ (illustrated in Figure~\ref{fig:triangle}). 
For a generic polygon $P$ with $n$ edges, we can inside-out dissect with only $2n+1$ pieces as follows. Take every external edge of $P$ and construct two isosceles triangles just as we did in the triangle example, and perform the same procedure, rotating and exchanging placing every external edge of $P$ in the interior (Figure~\ref{fig:adi-generic}).

\begin{lemma}
    Any regular polygon $P$ can be inside-out dissected with at most $6$ pieces.
\end{lemma}
\emph{Proof Sketch.} 
For $n = 3,4, \text{ and } 5$ we have ad hoc constructions (Figure~\ref{fig:inside-out-dissectionTriangle} for $n = 3$).
For $n \geq 6$, the rough idea is
to cut large rotationally symmetric pieces from $P$ and rotate them so perimeter edges become internal.
This is an extension of the isosceles triangles from Lemma~\ref{lem:triangle}. 

\begin{figure}[!htb]
    \centering
    \includegraphics[width=0.6\linewidth]{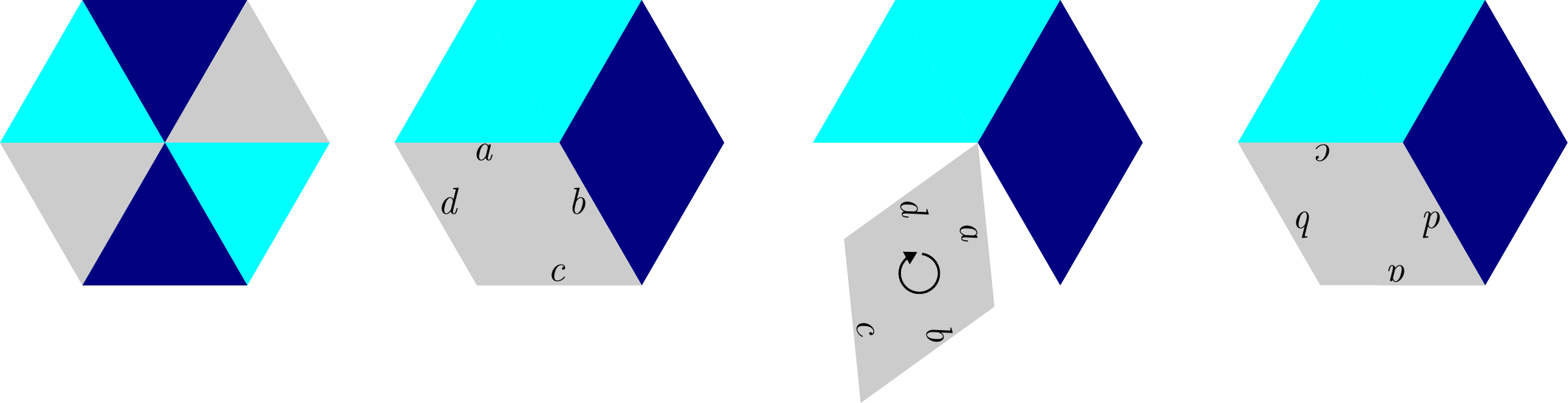}
    \caption{An inside-out dissection of a regular hexagon.}
    \label{fig:hexagon}
\end{figure}

\begin{figure}[ht]
    \centering
    \includegraphics[width=0.3\linewidth]{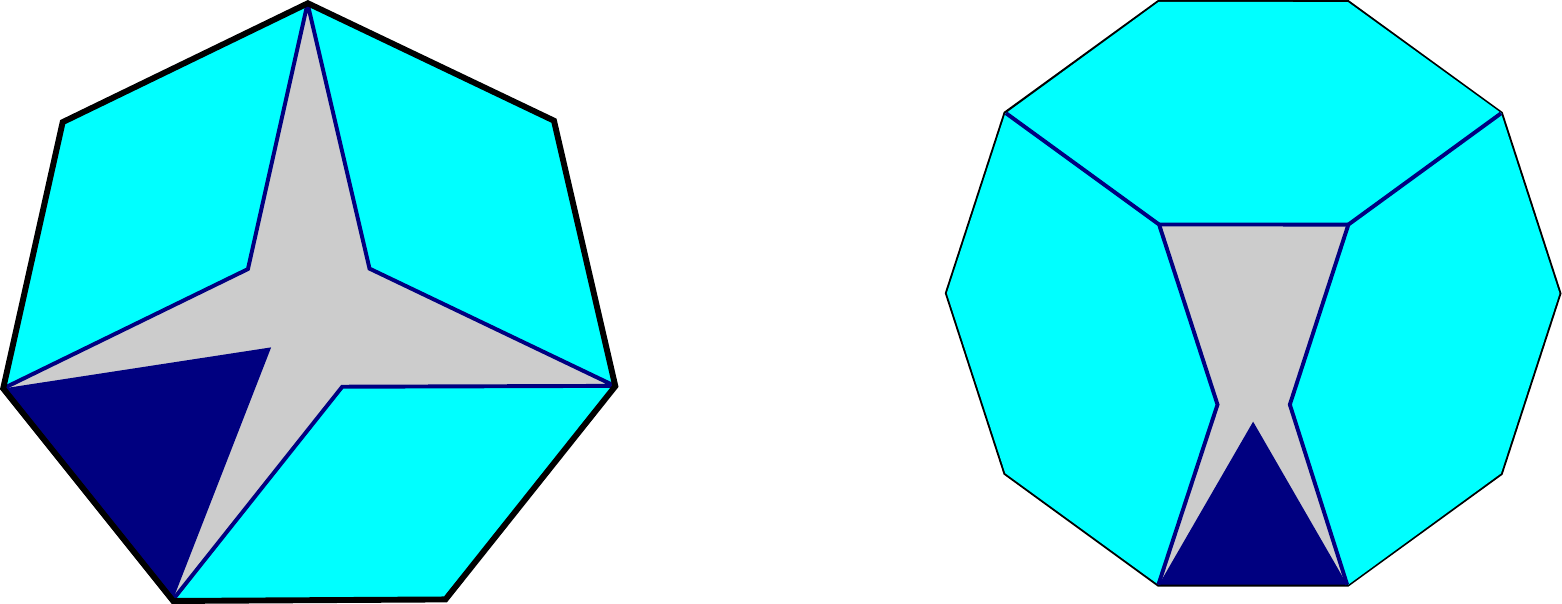}
    \caption{Two regular polygons with their pieces (left) 7-gon, (right) 10-gon.}
    \label{fig:generic-Regular}
\end{figure}

For the regular polygon $P$, each internal angle is $\frac{180(n-2)}{n}$ degrees. So, for $n = 6$, every internal angle is $\frac{180(6-2)}{6} = 120$ degrees. We can then cut a hexagon into three congruent rhombi by bisecting every other angle. Each rhombus can now be rotated 180 degrees inside-out dissecting a regular hexagon with three pieces, demonstrated in Figure~\ref{fig:hexagon}. 
For any regular polygon $P$, one can use the internal angle to compute the maximum rotationally symmetric $k$-gon that can be cut out from $P$.  With a detailed analysis of the angles of the given regular polygon $P$ we can show that at most $5$ such pieces will suffice to cover the perimeter of $P$. These pieces may not cover the area of $P$ and leave some piece of $P$ unaltered. Therefore, this method uses at most $5+1$ pieces (examples in Figure~\ref{fig:generic-Regular}).

\section{Dissections of Polyhedra}
 The main result of this section is Theorem \ref{theorem:polyhedra}, which establishes that polyhedra that admit a decomposition into a finite number of regular tetrahedra and octahedra can be inside-out dissected. To establish the desired result, we use the tetrahedral-octahedral honeycomb.

The \emph{tetrahedral-octahedral honeycomb} is the semi-periodic space-filling of the Euclidean $3$-space that consists of regular octahedra and tetrahedra with all edge lengths being $\sqrt{2}.$ 
The vertices of the tetrahedra and octahedra yielding this honeycomb can be described as points of the \emph{face-centred~cubic~lattice}, i.e.\ the three-dimensional lattice $\mathcal{L}$ given by 
$\mathcal{L}:=\langle v_1:=e_1+e_2,v_2:=e_1+e_3,v_3:=e_2+e_3\rangle,$
where $\{e_1,e_2,e_3\}$ denotes the standard basis of the real 3-space. For instance, by using this lattice we can construct a regular tetrahedron and a regular octahedron via $\conv(\{(0,0,0)^t,v_1,v_2,v_3\})$ and $\conv(\{v_1,v_2,v_3,v_1+v_2,v_1+v_3,v_2+v_3\})$, respectively.
We refer the reader to~\citep{symmetry} for a more general introduction to this honeycomb.

\begin{figure}[H]
    \centering
    \begin{tabularx}{\linewidth}{CC}
        \includegraphics[scale=0.25,page=1]{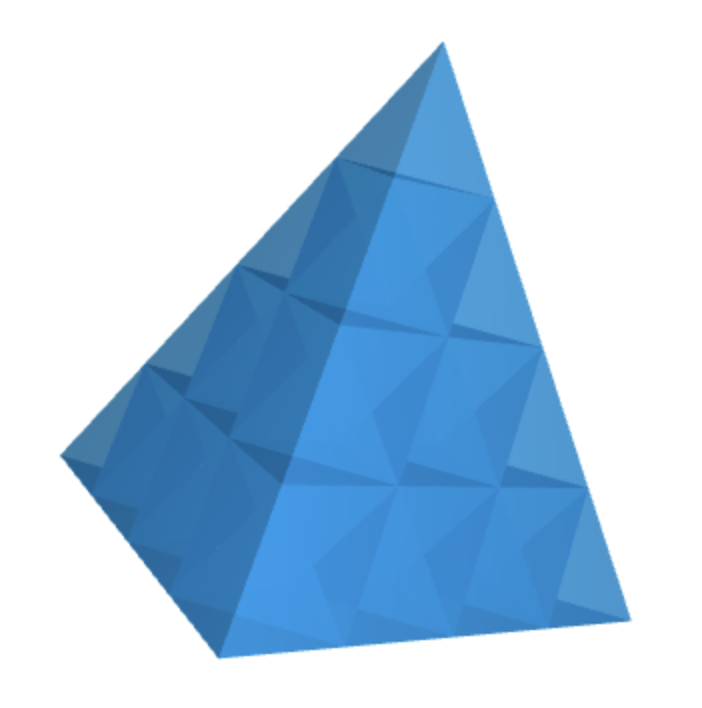}&%
        \includegraphics[scale=0.25,page=2]{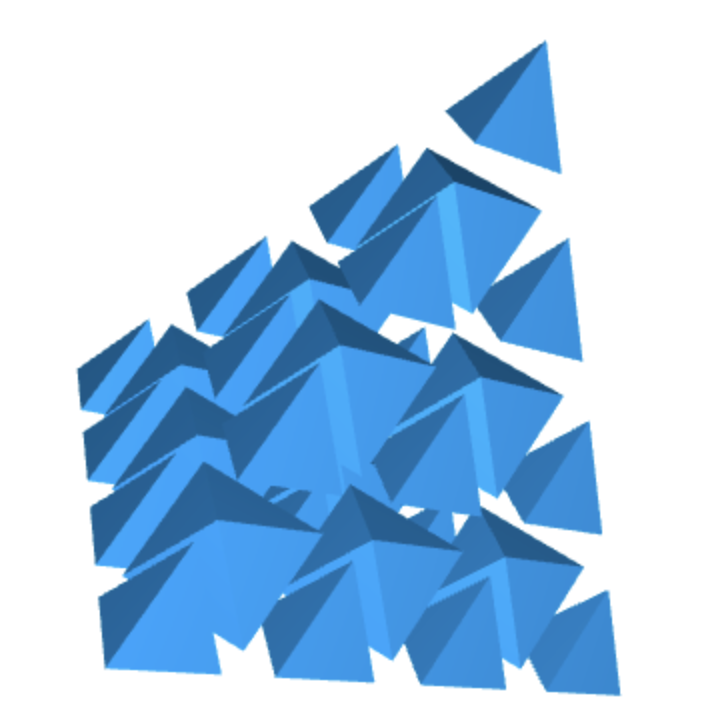}\\%
        (a) & (b)%
    \end{tabularx}
    \caption{(a) Subdivision of a regular tetrahedron with edge length $\sqrt{2}$ into $24$ regular tetrahedra and $10$ regular octahedra with all edges having length $\tfrac{\sqrt{2}}{4}$ (b) an exploded view of the described subdivision.}
    \label{fig:tetexp}
\end{figure}

It is well known that the structure of the tetrahedral-octahedral honeycomb can be exploited to decompose regular tetrahedra and octahedra into a finite number of smaller tetrahedra and octahedra.
\begin{remark}\label{remark:decom}
    Let $\T$ be the regular tetrahedron that is given by the convex hull $\conv (\{(0,0,0)^t,v_1,v_2,v_3\})$. It can be observed that $\T$ can be decomposed into $4$ regular tetrahedra and $1$ regular octahedron with all edge lengths being $\tfrac{\sqrt{2}}{2}.$ For $i=1,2,3$ we therefore define the vector $w_i$ as $w_i:=\tfrac{1}{2}v_i$. Thus, the desired decomposition of $\T$ ( see Figure~\ref{fig:insideOutTetrahedron}(a) ) is given by 
   $\T=\T_1\cup \T_2\cup T_3\cup T_4\cup \Oct,$
   where the polyhedra $T_1,\ldots,\T_4,\Oct$ are defined as
   \begin{align*}
      & \T_1:=\conv\{(0,0,0)^t,w_1,w_2,w_3\},       \T_2:=\conv\{w_1,w_1+w_2,w_1+w_3,2w_1\},\\
      & \T_3:=\conv\{w_2,w_1+w_2,w_2+w_3,2w_2\}
       \T_4:=\conv\{w_3,w_1+w_3,w_2+w_3,2w_3\},\\
       &\Oct:=\conv(\{w_1,w_2,w_3,w_1+w_2,w_1+w_3,w_2+w_3\}).
   \end{align*}
\begin{figure}[H]
    \centering
    \begin{tabularx}{\linewidth}{CC}
        \includegraphics[scale=0.1,page=1]{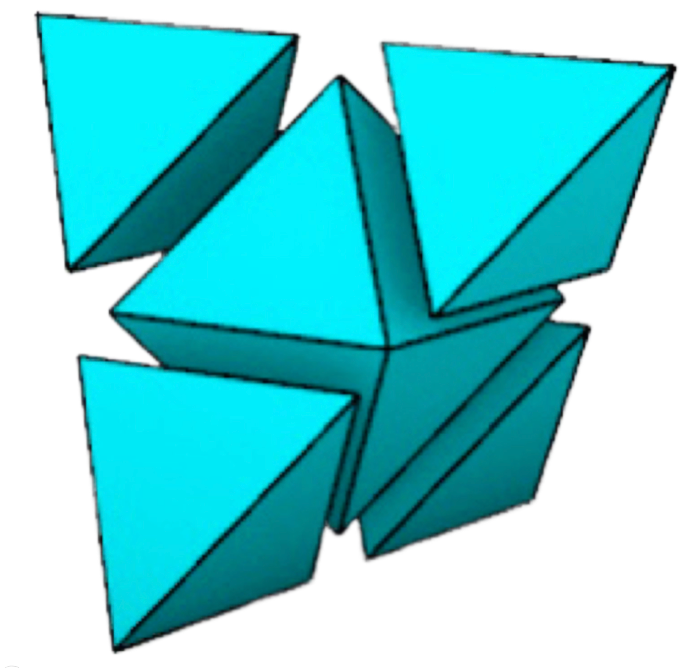}&%
        \includegraphics[scale=0.1,page=2]{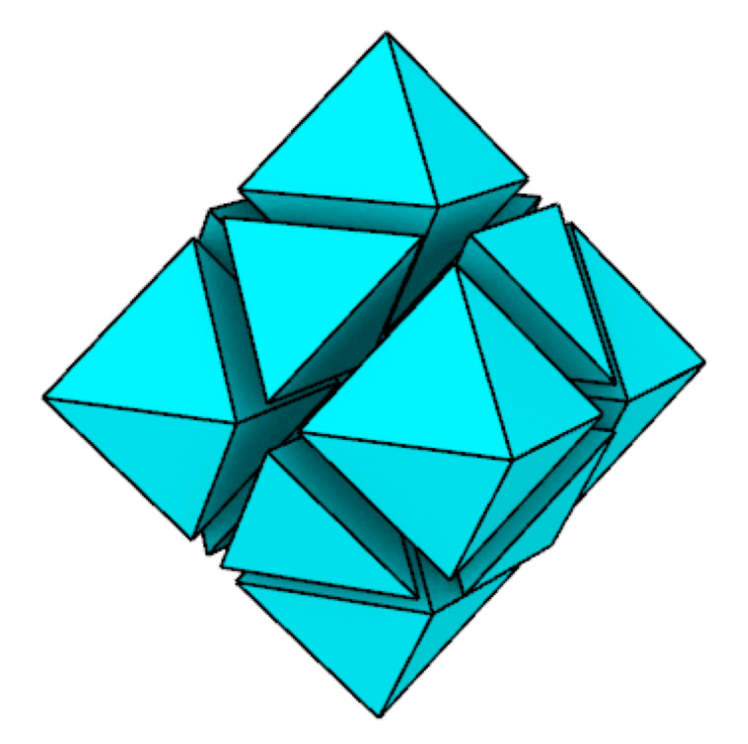}\\%
        (a) & (b)%
    \end{tabularx}
    \caption{Subdivision of (a) a regular tetrahedron into $4$ regular tetrahedra and $1$ regular octahedron and (b) a regular octahedron into $8$ regular tetrahedra and $6$ regular octahedra.~\cite{octtet}}
\label{fig:insideOutTetrahedron}
\end{figure} 
Similarly, by exploiting the vectors $w_1,w_2,w_3$ we can construct a decomposition of a regular octahedron with all edge lengths being $\sqrt{2}$ into $8$ regular tetrahedra and $6$ octahedra with all edge lengths being $\tfrac{\sqrt{2}}{2}$, (illustrated in Figure \ref{fig:insideOutTetrahedron}(b)).
\end{remark}
Using Remark~\ref{remark:decom}, we prove the following result.
\begin{lemma}\label{lemma:tet}
Let $\T$ be a regular tetrahedron. Then $\T$ can be inside-out dissected with $34$ pieces.
\end{lemma}
\emph{Proof Sketch.}
Since the tetrahedron $\T$ can be scaled such that its edge lengths are all equal to $\sqrt{2}$, we can assume that $\T$ is given by the convex hull $\conv(\{(0,0,0)^t,v_1,v_2,v_3\})$. By Remark~\ref{remark:decom} we can decompose $\T$ into $4$ regular tetrahedra and $1$ regular octahedron with all edge lengths being $\tfrac{\sqrt{2}}{2}.$ Now, again by using Remark~\ref{remark:decom} we can decompose the resulting tetrahedra and the resulting octahedron to obtain a decomposition of the tetrahedron $\T$ into $4\cdot 4+8=24$ regular tetrahedra and $4\cdot 1+1\cdot 6=10$ regular octahedra with all edge lengths being $\tfrac{\sqrt{2}}{4}.$ This decomposition is illustrated in Figure~\ref{fig:tetexp}.
We refer to the face of a polyhedron of the above decomposition as boundary face if it is contained in the boundary of $\T$.
Thus, it can be observed that the above decomposition has exactly 
$2\cdot 6+4+4=20$ tetrahedra and $10$ octahedra that have at least one boundary face. More precisely, there are exactly (1) $4$ tetrahedra having exactly $3$ boundary faces, (2) $12$ tetrahedra having exactly $2$ boundary faces, (3) $4$ tetrahedra having exactly $1$ boundary face, (4) $4$ octahedra having exactly $3$ boundary faces and (5) $6$ octahedra having exactly $2$ boundary faces. In order to construct the desired inside-out dissection of the tetrahedron $\T$ the tetrahedra described in (1) and (3) have to be swapped by applying a suitable transformation and the polyhedra in (2), (4) and (5) have to be rotated so that the boundary faces of these polyhedra do not form boundary faces of the rearranged tetrahedron.  
\vspace{0.3cm}

With similar arguments, the below result follows. 
\begin{lemma}\label{lemma:oct}
    Let $\Oct$ be a regular octahedron. Then $\Oct$ can be inside-out dissected with $124$ pieces.
\end{lemma}
We combine these two Lemma \ref{lemma:tet} and Lemma \ref{lemma:oct} the desired theorem is established.
\begin{theorem}\label{theorem:polyhedra}
    Every polyhedron $P$ that can be decomposed into a finite number of regular octahedra and tetrahedra can be inside-out dissected.
\end{theorem}
\section{Conclusion \& Outlook}
This manuscript leaves open several problems:
\begin{itemize}
\item Can a triangle be inside-out dissected with 3 pieces?
    \item Can general $n$-gons be done with less than $2n + 1$ pieces? Maybe convex polygons can be done with constant pieces like regular polygons?
    \item Can the presented method to inside-out dissect a regular tetrahedron and a regular octahedron be improved? Yes, we can show that a regular tetrahedron can be inside-out dissected with $13$ and a regular octahedron can be inside-out dissected with $25$ pieces.
    \item We have a method for general polyhedra, but the number of pieces required is quite large. Therefore, an efficient method for inside-out dissections of polyhedra is still open. 
\end{itemize}
\newpage
\bibliographystyle{plainnat}
\nocite{*}
\bibliography{main}
\end{document}